\pdfoutput=1
\documentclass[a4paper]{jpconf}
\usepackage{amsmath}
\usepackage{amssymb}
\usepackage{graphicx, color}
\begin{document}
\title{Squeezed coherent states and the one-dimensional Morse quantum system}

\author{M Angelova$^1$\ A Hertz$^2$ and V Hussin$^3$}

\address{$^1$Mathematical Modelling Lab, School of Computing,
Engineering and Information Sciences, Northumbria University,
Newcastle NE2 1XE, UK}
\address{$^2$D\'epartement de Physique,
Universit\'{e} de Montr\'{e}al, Montr\'{e}al, Qu\'ebec, H3C 3J7, Canada}
\address{$^3$D\'epartement de Math\'ematiques et de
Statistique,
Universit\'{e} de Montr\'{e}al, Montr\'{e}al, Qu\'ebec, H3C 3J7, Canada}

\ead{maia.angelova@northumbria.ac.uk, anaelle.hertz@umontreal.ca, veronique.hussin@umontreal.ca}

\begin{abstract}
The Morse potential one-dimensional quantum system is a realistic model for studying vibrations of atoms in a diatomic molecule. This system is very close to the harmonic oscillator one. We thus propose a construction of squeezed coherent states similar to the one of harmonic oscillator using ladder operators. Properties of these states are analysed with respect to the localization in position, minimal Heisenberg uncertainty relation, the statistical properties and illustrated with examples using the finite number of states in a well-known diatomic molecule.

\end{abstract}

\section{Introduction}

Coherent and squeezed states are known to be very important in many fields of physics. Coherent states were discovered in 1926 by Schr{\"{o}}dinger \cite{schro 26}, while squeezed states were introduced by Kennard in 1927 \cite{kenn 27}. However, these works were, in the main, ignored or forgotten until the sixties, when these states became very popular and received a lot of attention from both fields, mathematics and physics. Among many important papers, let us mention the works of Glauber \cite{Glauber}, Klauder  \cite{Klauder, KS}, and Nieto  \cite{Nieto}. In the particular field of quantum optics, the books of  Walls and Milburn \cite{Walls}, Gazeau \cite {Gazeau} and Rand   \cite{Rand} are very good reading which also consider the applications. The study of squeezed states for systems admitting an infinite discrete spectrum, obtained as a generalisation of coherent states, has been recently the center of much attention (see, for example, \cite{Braunstein, Hillery, Bergou, Sasaki1, Alvarez}). 

In modern developments, coherent states (CS) are standardly defined in three equivalent ways: displacement operator method, ladder (annihilation) operator method and minimum uncertainty method (for review see for example \cite{Nieto}). Initially defined for the case of the harmonic oscillator, coherent states have been generalised for other systems.  We can use, for example, the 
definition of Klauder \cite{Klauder01} saying that they are obtained as the following superposition of energy eigenstates $\{| \psi_{n}\rangle, n\in {\mathbb{N}}\}$
\begin{equation}
\psi(z) ={\frac{1}{\sqrt{{\cal N} (|z|^{2})}}}\sum_{n \in I} \frac{z^{n}}{\sqrt{\rho(n)}}| \psi_{n}\rangle.
 \label{genecs}
\end{equation}
The sum is taken over all the discrete values of $n$ and the set $I$ is usually infinite. The parameter $z$ is  a complex variable in general, ${\cal N}$ is a normalization factor and $\{\rho(n), n\in {\mathbb{N}}\}$ is a set of strictly positive parameters, usually depending on the energy of the system under consideration. These last quantities correspond to a moment problem (see \cite{Klauder01} for details).

For a quantum system which admits an  infinite discrete spectrum $\{| \psi_{n}\rangle, n=0,1,...\}$ and ladder operators $A^-$ and  $A^+$ acting on the
energy eigenstates as
\begin{equation}
A^-| \psi_{n}\rangle=\sqrt{k(n)}\ | \psi_{n-1}\rangle, \ A^+ | \psi_{n}\rangle=\sqrt{k(n+1)}\ | \psi_{n+1}\rangle,
\label{ladder0}
\end{equation}
these coherent states are defined as eigenstates of $A^-$. We thus get  \cite{Klauder01}
\begin{equation}
\rho(n)=\prod_{i=1}^{n} k(i), \quad \rho(0)=1.
\label{rho}
\end{equation}
Note that the quantity $k(i)$ is not unique and can be chosen to impose additional constraints to the ladder operators. In particular, for the harmonic oscillator, we have  $k(i)=i$,
\begin{equation}
[A^-,A^+] = \mathbb{I}, \quad H_{ho}=\hbar \omega (A^+ A^- +{1\over 2}),
\end{equation}
and the expression (\ref{genecs}) gives the usual coherent states.

Now, for a quantum system which admits a finite discrete spectrum like
the one which involves the Morse potential, various constructions of coherent
states have been adapted  \cite{Dong1, Roy, Recamier, Daoud}.
In a recent paper \cite{Angelova}, we have used ladder operators
\cite{Frank, Dong2} to construct different types of coherent states of the Morse
potential and have compared them with the so-called Gaussian coherent
states \cite{Fox}. In particular, such a construction has been
inspired by the approach mentioned above (see formula (\ref{genecs})) but where the set $I$ of values of $n$ is now finite. The coherent states are not exactly eigenstates of the
annihilation operator $A^-$ but we have shown \cite{Angelova} that,
in practice, the last terms on the right hand side of the sum in (\ref{genecs}) does not
contribute significantly. In some approaches (see, for example,
\cite{Draganescu}) these states are called pseudo-coherent states.

To our knowledge, squeezed coherent states for the Morse potential
have not been constructed. The aim of this paper is thus to
show that such a construction can be closely related to
the one for infinite spectrum systems. In fact, these states would be almost eigenstates of a
linear combination of the ladder operators. Moreover, we demonstrate using numerical evaluations that one set of these states, the energy-like, behave properly as squeezed states. 

In Section 2 we give a review of relevant results on squeezed coherent states and minimal uncertainty relations for a quantum system with infinite spectrum. In Section 3, starting with the definition of the Morse model and its ladder operators, we define the corresponding squeezed coherent states. We thus get two types of the states called oscillator-like and energy-like. In Section 4, we show that the energy-like states have stable trajectories in the phase space and minimize the uncertainty relation. Due to the complexity of the expressions of those states, numerical calculations are used. We end the paper with conclusions in Section 5.

\section{Squeezed coherent states for a quantum system with infinite spectrum}

As in the case of  harmonic oscillator, general squeezed coherent states  \cite{Nieto}, for a quantum system with an infinite discrete energy spectrum, may be constructed as the solutions of the eigenvalue equation:
 \begin{equation}
(A^-+\gamma A^+)\psi(z,\gamma)=z \ \psi(z,\gamma), \quad z,\gamma \in {\mathbb C}.
\label{eigenequA1}
\end{equation}
The mixing of $A^-$ and $A^+$ is said to be controlled by a squeezing parameter $\gamma$ and $z$ is called the coherent parameter. The coherent states are special solutions when $\gamma=0$. Conditions on $\gamma$ must be imposed for the states to be normalisable.

Squeezed coherent states (SCS) based on $su(2)$ or  $su(1,1)$ algebras \cite{Bergou, Sasaki1} and also direct sums of these algebras with the algebra $h(2)$ \cite {Alvarez}, have been constructed using group theoretical methods. This involves, in particular, the operators displacement  $D$ and squeezing $S$ similar to the ones of the harmonic oscillator. In fact, for $su(2)$ or  $su(1,1)$ algebras, $k(n)$ is a quadratic function of $n$.

More generally, equation (\ref{eigenequA1}) may be solved by using a direct expansion of $\psi(z,\gamma)$ in the form
\begin{equation}
\psi(z,\gamma)={\frac{1}{\sqrt{{\cal N}_{g} (z, \gamma)}}}\sum_{n=0}^{\infty}\frac{Z(z,\gamma,n)}{\sqrt{\rho(n)}}| \psi_{n}\rangle,
\label{kfact}
 \end{equation}
 with
  \begin{equation}
{\cal N}_{g} (z, \gamma)=\sum_{n=0}^{\infty}{{|Z(z,\gamma,n)|^2}\over{\rho(n)}},
 \end{equation}
 where $\rho(n)$ is given by (\ref{rho}).
Indeed, for the case $\gamma\neq 0$, inserting (\ref{kfact}) into  (\ref{eigenequA1}), we get  a 3-term recurrence relation
 \begin{equation}
Z(z,\gamma,n+1)- z \ Z(z,\gamma,n) +\gamma\  k(n) \ Z(z,\gamma,n-1)=0,  \ n=1,2,...
\label{recurrel}
\end{equation}
and without restriction, we take $Z(z,\gamma,0)=1$ and thus  $Z(z,\gamma,1)=z$. 

For the harmonic oscillator, $\rho(n)=n!$ and we get explicitly \cite{Alvarez,Yuen}:
\begin{equation}
Z_{ho}(z,\gamma,n)=\sum_{i=0}^{[\frac{n}{2}]}{\frac{n!}{i! (n - 2 i)!}} \left( -\frac{\gamma}{2}\right) ^i {z^{(n - 2 i)}}=\left(\frac{\gamma}{2}\right) ^{\frac{n}{2}} \mathcal{H} \left( n,{\frac{z}{\sqrt{2\gamma}}}\right) .
\label{SCSho}
 \end{equation}
 It is well-known that $| \gamma|<1$ for the states to be normalizable in this case.
In (\ref{SCSho}), we see that the Hermite polynomials $\mathcal{H}(n, w)$ take values on ${\mathbb C}$. These polynomials have interesting properties in terms of orthogonality, measure and resolution of the identity \cite{Szafraniec}.

For the harmonic oscillator, these states minimize the Schr\"{o}dinger-Robertson uncertainty relation \cite{Merzbacher}  which becomes the usual Heisenberg uncertainty relation for $\gamma$ real. 
Indeed, we get
\begin{equation}
{(\Delta \hat{x})}^{2}=  \frac{1}{1+ \gamma}-\frac{1}{2},\ {(\Delta \hat{p})}^2  =  \frac{1}{1- \gamma}-\frac{1}{2}
\label{dispersionxp}
\end{equation}
and the uncertainty becomes
\begin{equation}
\Delta (z, \gamma )={(\Delta \hat{x})}^2{(\Delta \hat{p})}^2 =\frac{1}{4}.
\label{dispersionxp}
\end{equation}
We see that this implies the reduction of the "quantum noise" on one of the 
observables while increasing it on the other. In the
following we will treat the case when the quantum noise is reduced
on the observable $x$ because we want a good localisation in the position. 

\section{The Morse potential and different types of squeezed coherent states\label{sec:ho}}

The Morse potential quantum system is a realistic model for studying vibrations of atoms in a diatomic molecule. Since this system is very close to the harmonic oscillator, the squeezed coherent states will be constructed following the procedure given for the harmonic oscillator, but we will deal with a finite number of eigenstates.

\subsection{The model}

The one-dimensional Morse model is given by the energy eigenvalue equation (see, for example, \cite{Frank}) 
\begin{equation}
{\hat{H}}\ \psi(x) =\left(\frac{{ \hat{p}}^2}{2m_r}+V_M(x)\right)\psi(x)= E\psi(x),
 \end {equation}
where $m_r$ is the reduced mass of the oscillating system composed of two
atoms of masses $m_1$ and $m_2$, {\it i.e.}
$\frac{1}{m_r}=\frac{1}{m_1}+\frac{1}{m_2}$. The potential is
$V_M(x)=V_0(e^{-2\beta x}-2e^{-\beta x})$, where  the space variable $x$ represents the displacement of the two atoms
from their equilibrium positions,  $V_0$ is a scaling energy
constant representing the depth of the potential well at equilibrium
$x=0$ and $\beta$ is the parameter of the model (related to the
characteristics of the well, such as its depth and width).

The finite discrete spectrum is known as
\begin{equation}
E_n=-\frac{ \hbar^2}{2m_r} \beta^2 \ {\epsilon_n}^2,
\label{energies}
\end{equation}
where
\begin{equation}
\epsilon_n=\frac{\nu-1}{2}-n=p-n, \quad  \nu=\sqrt{\frac{8{m_r}V_0}{ \hbar^2 \beta^2}},
\label{epsilonnu}
\end{equation}
and $\{n=0,1,2,...,[p]\}$,  with $[p]$ the integer part of
$p=\frac{\nu-1}{2}$. 
The following shifted energies
\begin{equation}
e(n)=\frac {2m_r}{ \hbar^2 \beta^2}
(E_n-E_0)=\epsilon_0^2-\epsilon_n^2=n(2p-n)
\label{shifteden}
\end{equation}
are useful for the construction of squeezed coherent states.
Using the change of variable $y=\nu e^{-\beta x}$,
we get the energy eigenfunctions, for the discrete spectrum, in terms of associated Laguerre polynomials, denoted by $L_{n}^{2\epsilon_n}$, as
\begin{equation}
\psi_n^\nu (x)= {\cal N}_n \ e^{-\frac{y}{2}} y^{\epsilon_n} L_{n}^{2\epsilon_n}(y),
\label{eigenfunMorse}
\end{equation}
where ${\cal N}_n$ is a normalization factor given by
\begin{equation}
{\cal N}_n= \sqrt{ \frac{\beta(\nu-2n-1)\Gamma(n+1)}{\Gamma(\nu-n)}}= \sqrt{ \frac{2\beta(p-n)\Gamma(n+1)}{\Gamma(2p-n+1)}}.
\label{normM}
\end{equation}
Since $p$ is related to physical parameters (see (\ref{epsilonnu})), it is not an integer in practice and ${\cal N}$ is never zero as expected. 

For many applications,
it is convenient to introduce the number operator $\hat{N}$ such that
\begin{equation}
\hat{N}\psi_n^\nu (x)= n \ \psi_n^\nu (x)
\end{equation}
and we see from (\ref{energies}) that $\hat{H}$ can be in fact written as
${\hat{H}}\ =-\frac{ \hbar^2}{2m_r} \beta^2 \ (p-\hat{N})^2 $.
\subsection{Ladder operators}

The ladder operators of the Morse system are defined as in (\ref{ladder0}), however the set of eigenfunctions $ \{| \psi_{n}\rangle\}$ is finite. As mentioned in the introduction, the quantity $k(n)$ in (\ref{ladder0}) is not unique and we consider two natural choices \cite{Angelova}.
The first choice, the "oscillator-like", corresponds to $k(n)=n$. The ladder operators satisfy a $h(2)$ algebra. The second choice, the "energy-like", corresponds to $k(n)=e(n)$, as given in (\ref{shifteden}). Ladder operators satisfy a $su(1,1)$ algebra.
In what follows, the subscripts {\it o} and
{\it e} will be used to refer to these choices.

Though our future calculations do not need the
explicit form of the ladder operators,  we give them for
completeness \cite{Daoud, Frank, Singh,
Sasaki}. For example, we get \cite{Frank}:
 \begin{eqnarray}
 A^-&=&-[\frac{d}{dy}(\nu-2N)-\frac{(\nu-2N-1)(\nu-2N)}{2 y}+\frac{\nu}{2}]\sqrt {K(N)}, \label{laddergen}\\
A^+&=& (\sqrt {K(N)})^{-1}[\frac{d}{dy}(\nu-2N-2)+\frac{(\nu-2N-1)(\nu-2N-2)}{2 y}-\frac{\nu}{2}] ,
\label{laddergen1}
   \end{eqnarray}
where $K(n)$ is related to $k(n)$ by
   \begin{equation}
   k(n)=\frac{n(\nu-n)(\nu-2n-1)}{\nu-2n+1}K(n).
   \end{equation}
These relations are  valid for any integer $n$ in the interval $[0,[p]-1]$. Note that, for $n=[p]$, we get a permitted energy eigenstate $\psi_{[p]}^\nu(x)$
of the Morse potential but the action of the creation operators on this state does not give zero in general.
It gives a state which may not be  normalisable with respect to our scalar product.
This problem has been already mentioned in some contributions (see, for example, \cite{Frank, Singh}).
For arbitrary $p$, the special choice $k(n)=n([p]+1-n)$, leads to $A_+ \psi_{[p]}^\nu(x)=0$ and  $A_+ \psi_{[p]-1}^\nu(x)=\sqrt{[p]}\psi_{[p]}^\nu(x)$. 

The "oscillator-like" ladder operators are thus obtained by taking 
\begin{equation}
K_{o}(n)=\frac{\nu-2n+1}{(\nu-n)(\nu-2n-1)},
\label{knoh}
\end{equation}
while we see that  $K_{e}(n)= K_{o}(n) (\nu-1-n)$ for the "energy-like" ladder operators.

 \subsection{The harmonic oscillator limit}
 
The harmonic oscillator limit \cite{Dong2} is obtained by first shifting the Morse potential by $V_0= \frac{k'}{2 \beta^2}$ to get
\begin{equation}
V_1=V_0(1-e^{-\beta x})^2=V_M+V_0,
\label{Morse1}
\end{equation}
and then taking $\beta \to 0$ so that
$V_1 \to \frac12 k' x^2$ where $k'$ is the force constant. Note that the new Hamiltonian with potential $V_1$ has thus the energy levels shifted and we get
\begin{equation}
E_n^1=- \frac{ \hbar^2}{2m_r} \beta^2 \left[ \left(\frac{\nu-1}{2}-n\right)^2- \left(\frac{\nu}{2}\right)^2\right].
\end{equation}
Since, $\nu$ is given by (\ref{epsilonnu}), we get here
$\nu=\frac{2 \sqrt{{m_r}k'}}{\beta^2 \hbar}$.
The oscillator limit is obtained when $\nu \to \infty$ giving, as expected, an infinite spectrum and the good limit for the energies
\begin{equation}
\lim_{\nu \to \infty} E_n^1= \hbar \sqrt{\frac{k'}{m_r}}\left( n+\frac12\right) .\nonumber
\end{equation}
Second, we have to take the limit on the ladder operators. We replace $\beta$ by its expression in terms of $\nu$ and define $c=\sqrt{\frac{4m_r k'}{\hbar^2}}$. The annihilation operator $A^-$, given in (\ref{laddergen}), thus takes the form:
\begin{equation}
A^-=  \sqrt {K(n)}\left[ \frac{e^{\sqrt{\frac{c}{\nu}} x}}{\sqrt{c\ \nu}}(\nu-2n) \frac{d}{dx}+ \frac{e^{\sqrt{\frac{c}{\nu}} x}}{2 \nu} (\nu-2n-1)(\nu-2n)-\frac{\nu}{2}\right] .
\end{equation}
Since $K(n)$ depends also on $\nu$, we have to take the limit carefully. To solve it, just take the Taylor expansion of the exponential to the first order. We then see that $K(n)$ must behave as ${\nu}^{-1}$, which is exactly what we get from (\ref{knoh}) and we find
\begin{equation}
\lim_{\nu \to \infty}  A^-= \frac{1}{\sqrt c}\left( \frac{d}{dx}+\frac{c}{2}x\right) .
\end{equation}
A similar calculation gives the expected limit for $A^+$.

\subsection{Squeezed coherent states and their time evolution }

The squeezed coherent states of the Morse Hamiltonian are now defined as the finite sum
\begin{equation}
\Psi^{\nu}(z, \gamma, x)=  \frac{1}{\sqrt{{\cal{N}}^{\nu} (z,\gamma)}}
\sum_{n=0}^{[p]-1} \frac{Z(z,\gamma,n)}{\sqrt{\rho(n)}} \psi_n^{\nu} (x),
\label{scsMorse}
\end{equation}
where $\rho(n)$ is given in (\ref{rho}), $Z(z,\gamma,n)$ satisfies (\ref{recurrel}) and
 \begin{equation}
{\cal N}^{\nu} (z, \gamma)=\sum_{n=0}^{[p]-1}{{|Z(z,\gamma,n)|^2}\over{\rho(n)}}.
 \end{equation}

Such a definition is relevant since we have seen in the preceding subsection that the "oscillator-like" ladder operators tend to the ones of the harmonic oscillator when $k(n)=n$ and the appropriate limit is taken. Moreover, these states are  "almost" eigenstates of a linear combination of the generic ladder operators $A^-$ and $A^+$ which can be written as:
\begin{equation}
(A^-+\gamma\ A^+) \ \Psi^{\nu}(z, \gamma, x)\approx z \Psi^{\nu}(z, \gamma, x).
\end{equation}
In fact, the correction can be computed using the recurrence relation (\ref{recurrel}) and we find
\begin{equation}
\chi^{\nu} (z,\gamma,[p],x)=\Lambda_{1}(z,\gamma,[p])\psi_{[p]-1}^{\nu} (x)+\Lambda_{0}(z,\gamma,[p]) \psi_{[p]}^{\nu} (x),
\end{equation}
where
\begin{eqnarray}
\Lambda_{1}(z,\gamma,[p])&=&\frac{1}{{\sqrt{\rho_{[p]-1}}}} Z(z,\gamma,[p]), \nonumber \\
\Lambda_{0}(z,\gamma,[p])&=&\frac{1}{{\sqrt{\rho_{[p]}}}}\gamma  k([p]) Z(z,\gamma,[p]-1).
\end{eqnarray}
In practice, the last two terms of the sum in (\ref{scsMorse}) have a very weak contribution which justifies thus the term  "almost" eigenstates used above.

Other constructions of squeezed coherent states have been considered (see, for example, \cite{Daoud, Dong2}). They implicitly use the displacement operator $D$. It must be questioned first because we are dealing with a finite number of eigenstates in (\ref{scsMorse}). Indeed, the action of this operator is not well defined even if we take a finite development of the exponentials. Moreover the Baker-Campbell-Hausdorff formulae for expanding $D$ (as products of exponentials of simple operators) is not necessarily valid (see, for example \cite{Sasaki}). Secondly, only one parameter is involved in this displacement operator, that is the reason why they are called coherent states by these authors  \cite{Daoud, Dong2} . They are, in fact, special cases of our squeezed coherent states where $z$ and $\gamma$ are not independent ($\gamma\neq 0$). 

Since our squeezed coherent states are closely related to the ones of the harmonic oscillator, we are interested in the behaviour of these states in the physical observable-position $x$ and observable-momentum $p$. Here these observables are not obtained as linear combinations of the ladder operators (as we can see from (\ref {laddergen}) and (\ref {laddergen1})) and we must compute the mean values explicitly. 

For an arbitrary observable $\cal \theta$, we have $\langle {\cal \theta}\rangle(z, \gamma;t)=\langle \Psi^{\nu}(z, \gamma, x;t)| {\cal \theta}|\Psi^{\nu}(z, \gamma, x;t)\rangle$, where
the time evolution of our squeezed coherent states is given by
 \begin{equation}
\Psi^{\nu}(z, \gamma, x;t)=  \frac{1}{\sqrt{{\cal{N}}^{\nu} (z, \gamma)}}
\sum_{n=0}^{[p]-1}  \frac{Z(z,\gamma,n)}{\sqrt{\rho(n)}} e^{- \frac{ i E_n}{\hbar}t} \psi_n^{\nu} (x)
\label{timestates}
\end{equation} 
and we get explicitly
   \begin{eqnarray}
\label{meanOex}
   \langle {\cal \theta}\rangle(z, \gamma;t) &=& \frac{1}{{\cal{N}}^{\nu} (z, \gamma)} \Biggl(
\sum_{n=0}^{[p]-1}  \frac{|Z(z,\gamma,n)|^2}{\rho(n)} \langle {\cal \theta}\rangle_{n,n}\nonumber \\
&+& \sum_{n=0}^{[p]-1}  \sum_{k=1}^{[p]-1-n}\frac{Z^*(z,\gamma,n+k)}{\sqrt{\rho(n+k)}}  \frac{Z(z,\gamma,n)}{\sqrt{\rho(n)}} \nonumber \\
&& (e^{- \frac{ i (E_{n+k}-E_n)}{\hbar}t} \langle {\cal \theta}\rangle_{n,n+k}+e^{ \frac{ i (E_{n+k}-E_n)}{\hbar}t} \langle {\cal \theta}\rangle_{n+k,n})\Biggr),
  \end{eqnarray}
  where
\begin{equation}\label{meanOmn}
   \langle {\cal \theta}\rangle_{m,n}=  \langle \psi_m^{\nu} | {\cal \theta}|\psi_n^{\nu} \rangle.
   \end{equation}

In the following developments,  we are considering  observables which are such that $\langle {\cal \theta}\rangle_{m,n}$ are symmetric or skewsymmetric with respect to the exchange of $m$ and $n$. We are thus led to two different cases. If  $\langle {\cal \theta}\rangle_{n+k,n}= \langle {\cal \theta}\rangle_{n,n+k}$, we get
    \begin{eqnarray}
\label{meanOsym}
    \langle {\cal \theta}\rangle(z, \gamma;t) &=& \frac{1}{{\cal{N}}^{\nu} (z, \gamma)} 
\sum_{n=0}^{[p]-1}  \frac{|Z(z,\gamma,n)|^2}{\rho(n)} \langle {\cal \theta}\rangle_{n,n}\nonumber \\
&+&  \frac{2}{{\cal{N}}^{\nu} (z, \gamma)}  \sum_{n=0}^{[p]-1}  \sum_{k=1}^{[p]-1-n}      \Biggl[      Re\Bigg( \frac{Z^*(z,\gamma,n)}{\sqrt{\rho(n)}}  \frac{Z(z,\gamma,n+k)}{\sqrt{\rho(n+k)}}\Bigg)
\cos( \alpha(n,k) t)\nonumber \\
 &+&   Im\Bigg( \frac{Z^*(z,\gamma,n)}{\sqrt{\rho(n)}}  \frac{Z(z,\gamma,n+k)}{\sqrt{\rho(n+k)}}\Bigg) \sin( \alpha(n,k) t)\Bigg]  \langle {\cal \theta}\rangle_{n+k,n},
  \end{eqnarray}
  where 
  \begin{equation}
\alpha(n,k)=\frac{\hbar \beta^2}{2 m_{r}} k(2(p-n)-k).
  \end{equation}
  
If  $\langle {\cal \theta}\rangle_{n+k,n}= -\langle {\cal \theta}\rangle_{n,n+k}$, we get
    \begin{eqnarray}
\label{meanOantisym}
    \langle {\cal \theta}\rangle(z, \gamma;t) &=&   \frac{2 i}{{\cal{N}}^{\nu} (z, \gamma)}  \sum_{n=0}^{[p]-1}  \sum_{k=1}^{[p]-1-n}      \Biggl[      Re\Bigg( \frac{Z^*(z,\gamma,n)}{\sqrt{\rho(n)}}  \frac{Z(z,\gamma,n+k)}{\sqrt{\rho(n+k)}}\Bigg)
\sin( \alpha(n,k) t)\nonumber \\
 &-&   Im\Bigg( \frac{Z^*(z,\gamma,n)}{\sqrt{\rho(n)}}  \frac{Z(z,\gamma,n+k)}{\sqrt{\rho(n+k)}}\Bigg) \cos( \alpha(n,k) t)\Bigg]  \langle {\cal \theta}\rangle_{n+k,n}.
  \end{eqnarray}
   
   We get, from \cite{Sage}, the mean values of $\hat{x}$ and $\hat{p}$ (with $\beta=1$)
   \begin{equation}
  \langle \hat{x}\rangle_{n+k,n}=(-1)^{k+1}{{\cal{N}}_{n+k}}{{\cal{N}}_{n}} \frac{\Gamma(\nu-k-n)}{k (\nu-k-1-2n)n!}, \ k\neq 0,
  \end{equation}
   \begin{equation}
  \langle \hat{x} \rangle_{n,n}=\ln \nu-\Phi(0,\nu-1-2n)+ \sum_{j=1}^n \frac{1}{\nu-n-j},
  \end{equation}
  where the function $\Phi(0,z)=\frac{d}{dz} \ln \Gamma (z)$ and
    \begin{equation}
  \langle \hat{p} \rangle_{n+k,n}=i \hbar (-1)^{k+1} {{\cal{N}}_{n+k}}{{\cal{N}}_{n}} \frac{\Gamma(\nu-k-n)}{2 \ n! } (1- \delta_{k0}).  
  \end{equation}
We also get after some calculations,
   \begin{equation}
  \langle \hat{p}^2\rangle_{n+k,n}=\hbar^2 (-1)^{k+1} {{\cal{N}}_{n+k}}{{\cal{N}}_{n}}\frac{\Gamma(\nu-k-n)}{ 4 \ n! } ((k-1) \nu-k(k+2n+1)), \ k\neq 0
  \end{equation}
  and
   \begin{equation}
  \langle \hat{p}^2 \rangle_{n,n}=-\hbar^2\frac{(2n+1)(2n+1-\nu)}{4} .
  \end{equation}
The computation of the mean values of $\hat{x}^2$ is more tricky since it involves the functions ${\rm \Phi}(0,z)$ and ${\rm \Phi}(1,z)$. We do not have an analytic expression but we will be able to compute explicitly $\langle \hat{x}^2\rangle_{n+k,n}$ and $\langle \hat{x}^2\rangle_{n,n}$ since we have a finite number of these expressions to plug in $\langle {\hat{x}^2}\rangle(z, \gamma;t)$.

\subsection{Oscillator-like squeezed coherent states }

In this case, in accordance with the expression of  $\rho(n)$ given in (\ref{rho}), we take $k(i)=i$ and $\rho(n)=n!$ and we get $Z_{o}(z,\gamma,n)=Z_{ho}(z,\gamma,n)=(\ref{SCSho})$. For the special case where $\gamma=0$, we get  $Z_{ho}(z,0,n)=z^n$ while, for the squeezed vacuum $z=0$, we get
\begin{equation}
Z_{o}(0,\gamma,2n)={\frac{(2n)!}{n!}} \left( -\frac{\gamma}{2}\right) ^n, \quad Z_{o}(0,\gamma,2n+1)=0.
\label{vacuumho}
 \end{equation}
Moreover, in those states, we get the same probability distribution as for the harmonic oscillator:
 \begin{equation}
P_{o}(z,\gamma, n)= |\bigl\langle\psi_{n}\nu( x) | \psi \left( z,\gamma, x\right) \bigr\rangle|^2={\frac{1}{{\cal N}_{o}^\nu (z, \gamma)}}\left( {\frac{|\gamma|}{2}}\right) ^n {{| \mathcal{H}(n,\frac{z}{\sqrt{2\gamma}})|^2}\over{ n!}}
\label{probaoh}
\end{equation}
with
\begin{equation}
{\cal N}_{o}^\nu (z, \gamma)=\sum_{n=0}^{[p]-1} \left( {\frac{|\gamma|}{2}} \right) ^n {{| \mathcal{H}(n,\frac{z}{\sqrt{2\gamma}})|^2}\over{n!}}.
\end{equation}
The mean value and dispersion of the number operator $\hat{N}$ are now given by
 \begin{equation}
\langle  \hat{N} \rangle_{o}=\sum_{n=0}^{[p]-1}n \ P_{o}(z,\gamma,n), \ (\Delta \hat{N})^2_{o}=\sum_{n=0}^{[p]-1}n^2 P_{o}(z,\gamma,n)-(\sum_{n=0}^{[p]-1}n\ P_{o}(z,\gamma,n))^2.
\label{enoh}
 \end{equation}
The statistical properties of these states are similar to the ones of the harmonic oscillator since we get essentially the same quantity for  the Mandel's $Q$-parameter \cite{Mandel} given in general by 
\begin{equation}
Q(z,\gamma)=  \frac{(\Delta \hat{N})^2-\langle  \hat{N} \rangle}{\langle  \hat{N} \rangle}.
\label{corr}
\end{equation}
The only difference is that, in the calculation of  the dispersion and mean values in $\hat{N}$, the sums are now finite.  In particular, it is  well-known (see, for example, \cite{Walls})
that the probability density is a Poisson distribution in the special coherent case
($\gamma=0$). 

\subsection{Energy-like squeezed coherent states }

In this case, in accordance with the expression of  $\rho(n)$ given in (\ref{rho}), we take $k(i)=i(2p-i)$ and $\rho(n)=(-1)^{n} n!  (1-2p)_n$ where  $(a)_n$ is the usual notation for the Pochhammer symbol
\begin{equation}
(a)_n= a(a+1)(a+2)...(a+n-1)=\frac{\Gamma(a+n)}{\Gamma(a)}.
\end{equation}
The exact (infinite) recurrence relation
(\ref{recurrel}) can be solved directly in terms of hypergeometric functions. The solution is
\begin{equation}
Z_{e}(z,\gamma,n)=(-1)^n \gamma^{\frac{n}{2}} \frac{\Gamma(2p)}{\Gamma(2p-n)}\ {}_2 F_1\left(\begin{matrix}-n, - {\frac{z}{2 \sqrt\gamma}}+{\frac{1-2p}{2}}  \\ 1 - 2p\end{matrix};2\right), \ n=1,2,...,[p]-1.
\label{Zenergy}
\end{equation}
Since this result is far from being trivial, we give some details of the proof and also the expressions of few first polynomials of this sequence.

Let us first set
 \begin{equation}
k(n)=n(A-n), \ A\in {\mathbb{R}},
\label{knquad}
\end{equation}
so that the recurrence relation (\ref{recurrel}) becomes
  \begin{equation}
Z(z,\gamma,n+1)- z \ Z(z,\gamma,n) +\gamma\ n(A-n) \ Z(z,\gamma,n-1)=0,  \ n=1,2,...
\label{recurrelk}
\end{equation}
with $Z(z,\gamma,0)=1$ and thus  $Z(z,\gamma,1)=z$. Since we know the solution for the harmonic oscillator (i.e. when $k(n)=n$), we follow the same lines to solve (\ref{recurrelk}) for an infinite sequence of values of $n$. We introduce the new complex variable $w=\frac{z}{\sqrt{2\gamma}}$ and we take
\begin{equation}
Z(z,\gamma,n)=\left( \frac{\gamma}{2} \right) ^{\frac{n}{2}} f(n, w),
\label{bigz}
\end{equation}
We thus get a new recurrence relation on the functions $f(n, w)$:
 \begin{equation}
f(n+1,w)- 2 w \ f(n,w) +2 n(A-n) \ f(n-1,w)=0, \ f(1,w)=2 w, \ f(0, w)=1, \ n=1,2,...
\label{recurrencef}
\end{equation}
It is easy to see that $f(n, w)$ is in fact a polynomial of degree $n$ in $w$. Moreover, it can be expressed in terms of hypergeometric functions of the type ${}_2 F_1$. We explicitly get
\begin{equation}
 f(n,w)= 2^{\frac{n}{2}} (-A+1)_n \  {}_2 F_1\left(\begin{matrix}-n, - {\frac{w}{\sqrt2}}+{\frac{1-A}{2}} \\1 - A\end{matrix};2\right)
\end{equation}
and the hypergeometric function is in fact a polynomial in $w$ since we have
\begin{equation}
{}_2 F_1\left(\begin{matrix}-n, -v\\1 - A\end{matrix};2\right)= \sum_{k=0}^{n} \frac{2^k}{k!} \frac{(-n)_k(-v)_k}{(-A+1)_k}.
\end{equation}

The original function (\ref{bigz}) thus takes the form (\ref{Zenergy}) when $A=2p$ as expected. It is valid for any real value of $A$ and in fact, we see that the first polynomials of the sequence are given by
\begin{eqnarray}
Z_{e}(z,\gamma,0)&=&1, \ Z(z,\gamma,1)=z, \nonumber \\
Z_{e}(z,\gamma,2)&=&z^2-(A-1)\gamma,\nonumber \\
Z_{e}(z,\gamma,3)&=&z^3 - (3A-5) \gamma z,\nonumber \\
Z_{e}(z,\gamma,4)&=&z^4 - 2 (3A-7) z^2 \gamma + 3 (A-1)(A-3) \gamma^2\nonumber.
\end{eqnarray}

To be complete, let us mention that for the special case where $A$ is an integer, we see that the recurrence relation  (\ref{recurrencef}) splits in two different ones. Indeed, we get first a finite sequence of $f(n,w)$ satisfying (\ref{recurrencef}) for $n=1,2, ..., A-1$
and, second an infinite sequence of $f(n,w)$ for $n=A,A+1, ...$ satisfying the recurrence relation
 \begin{equation}
f(A+k+1,w)- 2 w \ f(A+k,w) -2k(A+k) \ f(A+k-1,w)=0, \ k=0,1,2,...
\label{recurrenceinf}
\end{equation}
Since for $k=0$, we get $f(A+1,w)=2w f(A,w)$, we can write $f(A+k,w)=h(k,w) f(A,w)$ where $h(k,w)$ is  a polynomial of degree $k$ in $w$ satisfying the recurrence relation
 \begin{equation}
h(k+1,w)- 2 w \ h(k,w) +2 k(-A-k) \ h(k-1,w)=0, \ h(1,w)=2 w, \ h(0, w)=1, \ k=1,2,...,
\label{recurrencek}
\end{equation}
which is (\ref{recurrencef}) where $A$ has been replaced by $-A$. The polynomials $h(k,w)$ are thus given by 
\begin{equation}
 h(k,w)= 2^{\frac{k}{2}} (A+1)_k \  {}_2 F_1\left(\begin{matrix}-k, - {\frac{w}{\sqrt2}}+{\frac{1+A}{2}} \\ 1+A\end{matrix};2\right),  \ k=0,1,2,...
\end{equation}
The solutions  $f(n,w)$ satisfying (\ref{recurrencef}) for $n=1,2, ..., A-1$ are in fact associated to a finite sequence of Krawtchouk polynomials while the solutions  $h(n,w)$ for $n=0,1,...$ are associated with Meixner polynomials \cite{Koe}. They both satisfy discrete orthogonality relations on the variable $w$ but these are not relevant in our context since $w$ is a continuous parameter.

When $\gamma=0$, we get  $Z_{e}(z,0,n)=z^n$ leading to coherent states while for the squeezed vacuum ($z=0$), we get
\begin{equation}
Z_{e}(0,\gamma,2n)=4^{(n-1)} (1 -2p) (3/2)_{n-1} (
  3/2 - p)_{n-1}\gamma^n, \quad Z_{e}(0,\gamma,2n+1)=0.
  \label{vacuume}
 \end{equation}

Now the probability distribution, denoted by $P_{e}(z,\gamma,n)_e$, is given by
\begin{equation}
P_e(z,\gamma, n)= {\frac{1}{{\cal N}_e^\nu (z, \gamma)}}\frac {\Gamma(2p-n) }{\Gamma(2p)n!}|Z(z,\gamma,n)|^2,
\label{pn}
\end{equation}
where
\begin{equation}
{\cal N}_e^\nu (z, \gamma)=\sum_{n=0}^{[p]-1} \frac {\Gamma(2p-n)}{\Gamma(2p)n!}|Z(z,\gamma,n)|^2.
\end{equation}
Similar expressions for $\langle  \hat{N} \rangle_e$ and $(\Delta \hat{N})^2_e$ are obtained as in (\ref{enoh}).

\section{Behaviour of the squeezed coherent states for the case of the hydrogen chloride molecule}

For the hydrogen chloride molecule $^1${\rm H}$^{35}${\rm Cl} (already considered in a previous paper \cite{Angelova}), we will fix the values of the physical parameter $\nu$ given in (\ref{epsilonnu}) with published values of $m_r$, $\beta$ and $V_0$, or as most often in practice, using $\nu = {{\omega}_e/{{\omega}_e x_e}}$, the ratio between the experimentally measured molecular harmonicity $\omega_e$ and anharmonicity $\omega_e x_e$ constants (see for example \cite{Herzberg50, Guela, CRC,AF07}). For the ground state, $X^1 \Sigma^+$ we have $\nu\approx57.44$ and $[p]=28$. We also choose the units such that $\frac{\hbar}{2 m_r}=1$ and $\beta=1$.

Let us start with some general facts. First, the squeezing parameter must be restricted to $|\gamma|<1$ for the states to be normalisable (as for the harmonic oscillator). 
Second, the minimisation of the Heisenberg uncertainty relation depends on the values of $z$ when $\gamma$ is real (and less than 1). Finally, we will see that the squeezing effect(reduction of the dispersion of one observable at the price of increasing it in the other) is always present even for $\gamma=0$. Moreover, the dispersion in $\hat{x}$ may be chosen smaller for the energy-like than for the oscillator-like states leading to a better localisation for the energy-like states. 

 In order to compare our results with the well-known ones for the harmonic oscillator, we restrict ourselves to positive real values of $z$ and $\gamma$ with $\gamma<1$. In fact, no dramatic difference appear when $z$ and/or $\gamma$ are negatives. 

\subsection{Coherent system of states ($\gamma=0$)}

\begin{figure}[h]
\centering
\includegraphics[width=.4\textwidth]{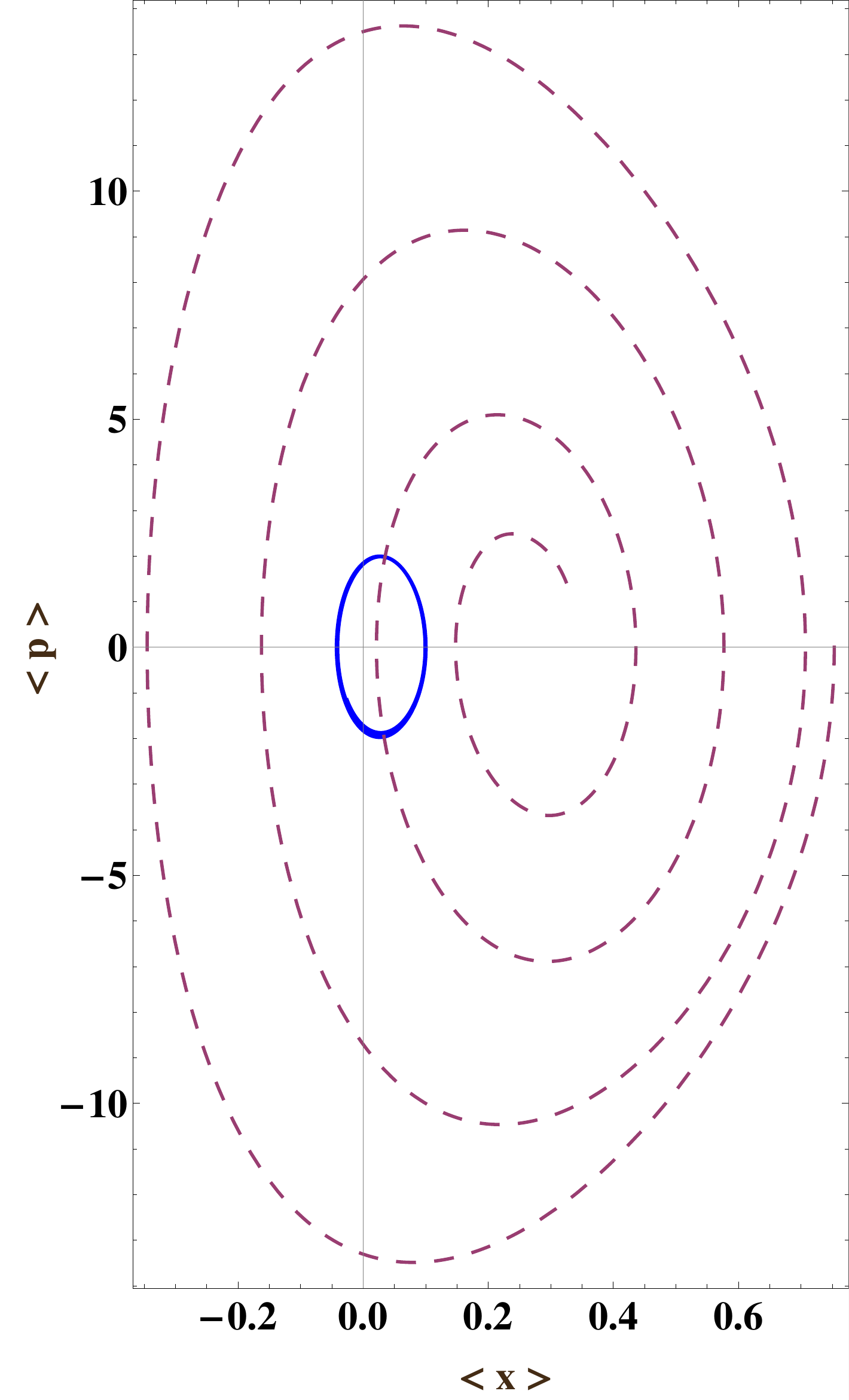}
\caption{ {\small Phase-space trajectories for oscillator-like (dashed line) and  energy-like (plain line) states when  $(z,\gamma )=(2,0)$ and $t\in[0,1]$.}}
\label{Fig1}
\end{figure}

The oscillator-like and energy-like states differ only by the denominator (and the normalisation factor)  in the development  \eqref{scsMorse}. In Fig \ref{Fig1}, we show the trajectories for both states with $z=2$ and $t\in[0,1]$ and we see that the oscillator-like states are less stable than the energy-like ones for the same value of $z$. This is true when $z<20$. We also see that, even if $\gamma=0$, squeezing always appears in the coherent system of states for the Morse potential.

\begin{figure}[h]
\centering
\includegraphics[width=.4\textwidth]{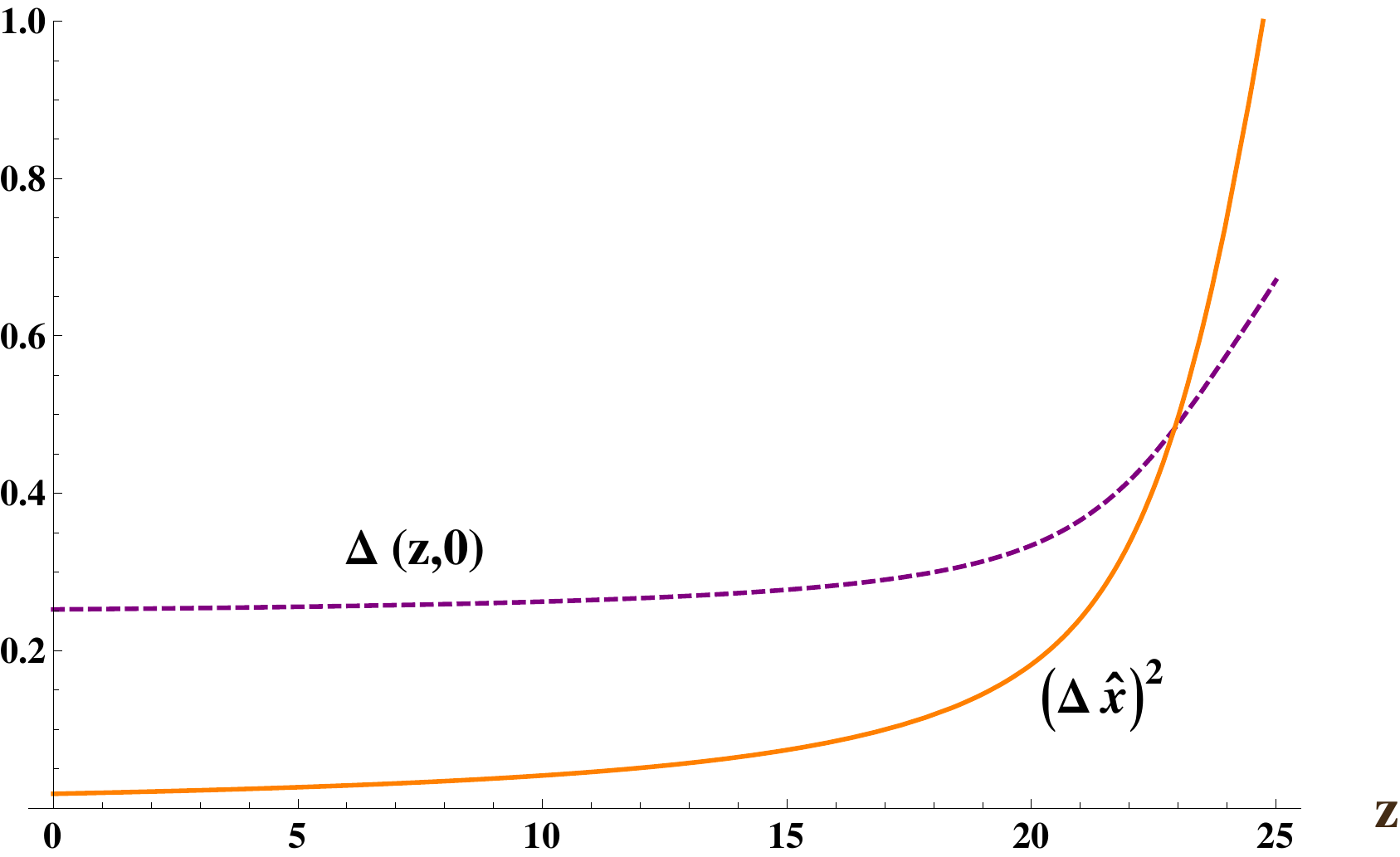} \qquad \qquad
\includegraphics[width=.45\textwidth]{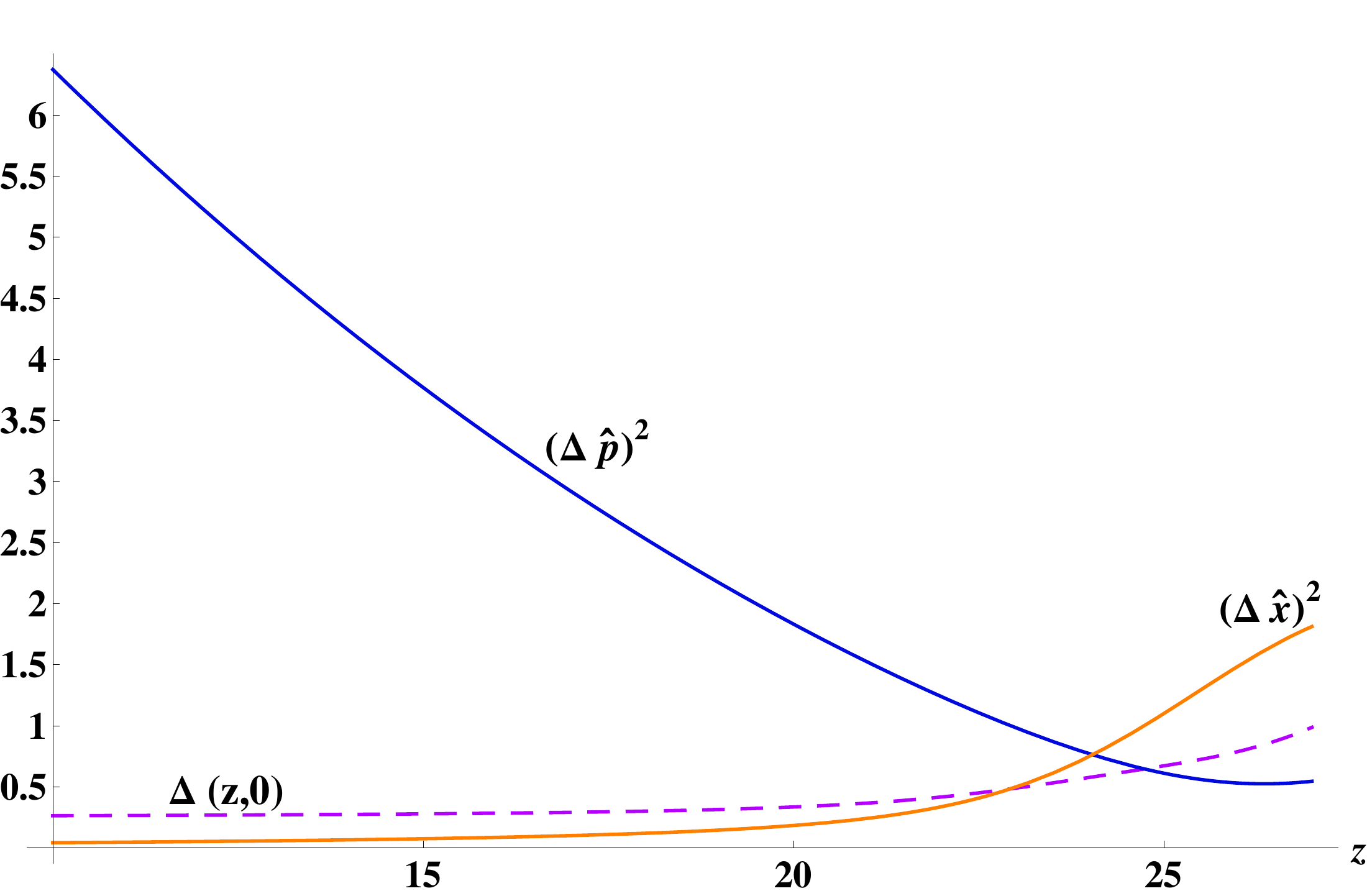}
\caption{ {\small Uncertainty and dispersions for the energy-like states at $\gamma=0$, $\Delta$ and $(\Delta \hat{x})^2$ with $z\in ]0,25]$ (left) and $\Delta$, $(\Delta \hat{x})^2$ and $(\Delta \hat{p})^2$ with $z\in [10,27]$ (right).}}
\label{Fig2}
\end{figure}

At $t=0$, the uncertainty product $\Delta(z, 0)=(\Delta x)^2 (\Delta p)^2$ and the dispersion $(\Delta x)^2$ show a very stable behaviour in  $z$ for the energy-like states (see  Fig \ref{Fig2}). These states are mostly minimal uncertainty states and the dispersion in $x$ is very small leading to a very good localisation (at least for $z<20$). This last fact is confirmed by the large eccentricity (in fact, it is very close to $1$) of the ellipses obtained before.
Interestingly, if we want to minimise the dispersion in $p$, we see, again on Fig \ref{Fig2}, that  $\Delta(z, 0)$ is no longer minimized. Also, reducing the dispersion in $p$ increases the dispersion in $x$. Moreover, we can easily see that the best minimisation of the dispersion in $p$ is observed in this special case, i.e. for $\gamma=0$. The oscillator-like states have almost the same behaviour but $\Delta(z, 0)$ increases faster for smaller values of $z$ (around $z=3$).
   
\begin{figure}[h]
\centering
\includegraphics[width=.4\textwidth]{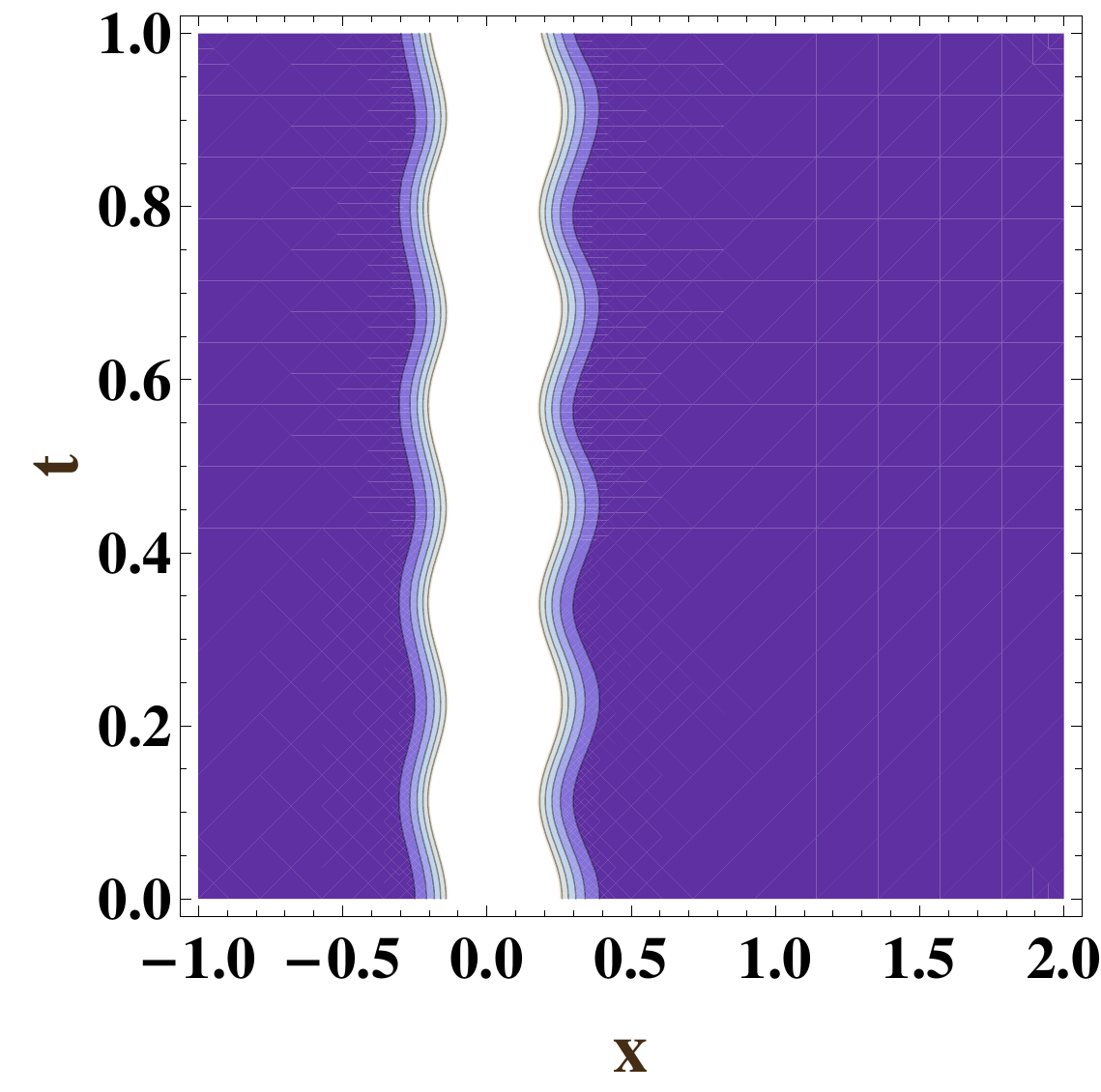}\qquad \qquad
\includegraphics[width=.4\textwidth]{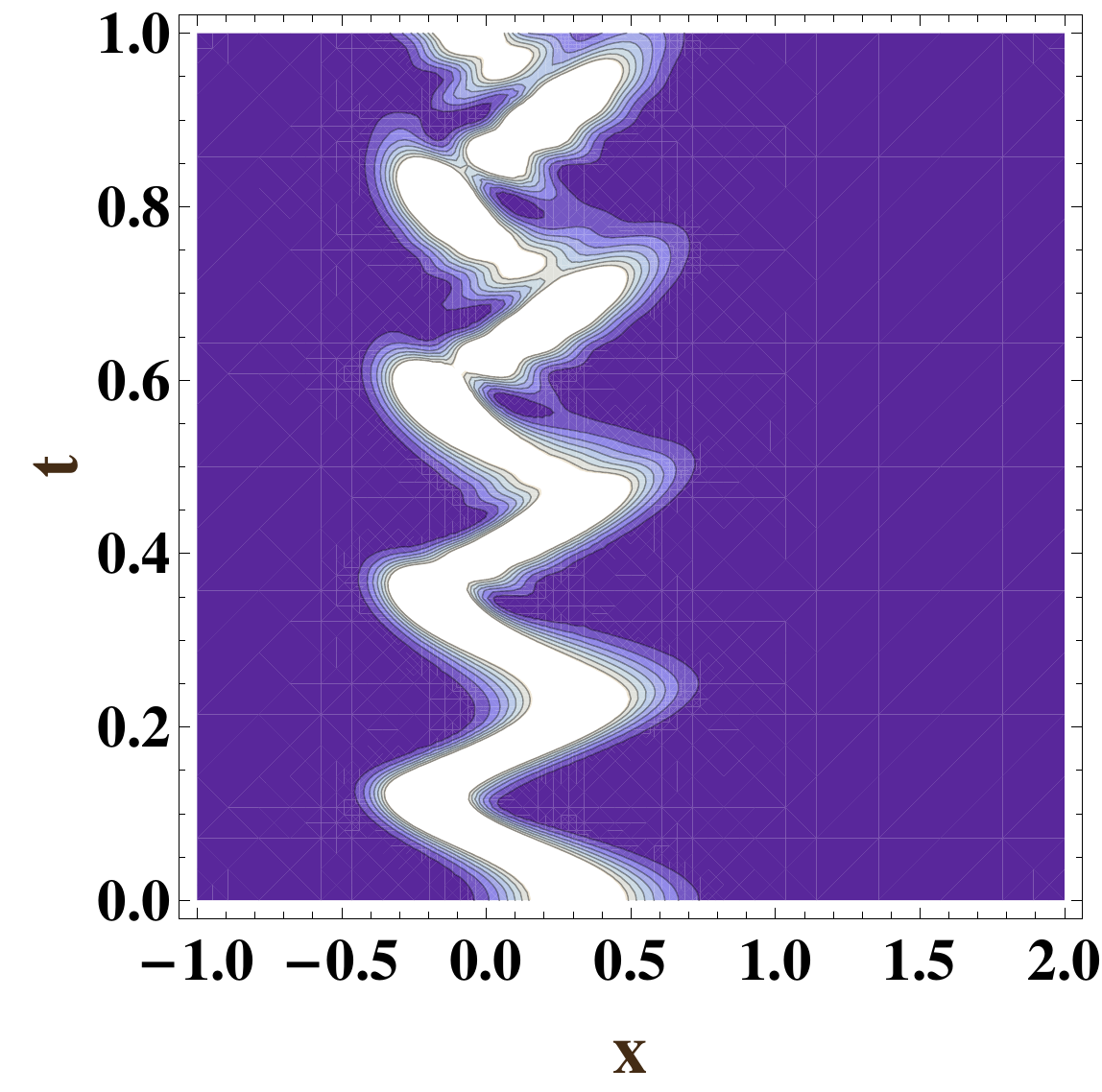}
\caption{ {\small Density probability $|\Psi^{\nu}_e(1, 0,x;t)|^2$ (left) and $|\Psi^{\nu}_o(1, 0,x;t)|^2$ (right) for $x\in[-1,2]$ and $t\in [0,1]$.}}
\label{Fig3}
\end{figure}

Finally, the density probabilities of both types of states have been computed to confirm the preceding results. In  Fig \ref{Fig3}, the energy-like states show again a better behaviour then the oscillator-like states as time evolves.

\subsection{Squeezed vacuum ($z=0$)}

\begin{figure}[!h]
\centering
\includegraphics[width=.4\textwidth]{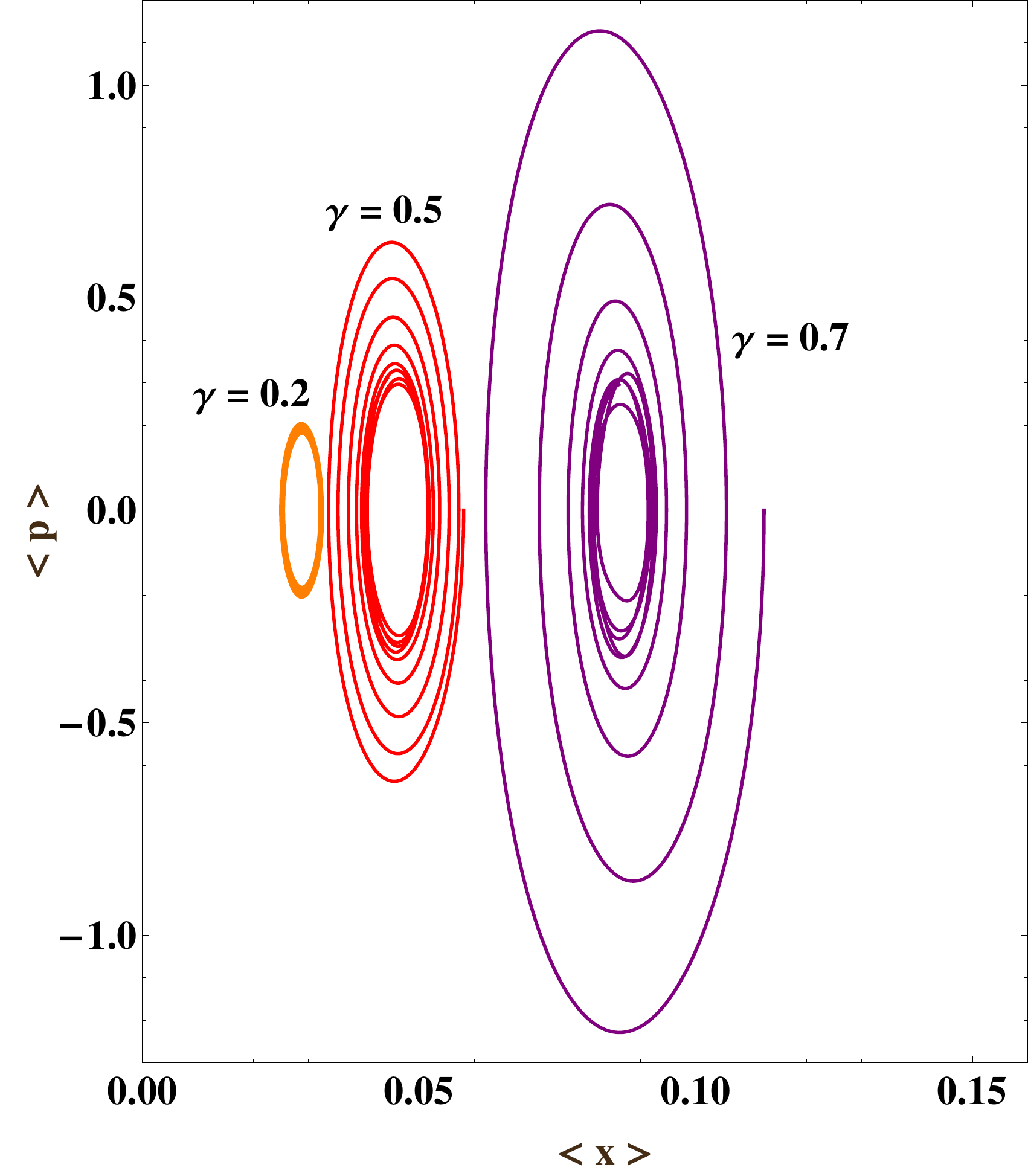}
\caption{ {\small Phase-space trajectories for energy-like states in the vacuum with $\gamma=0.2,0.5,0.7$ and $t\in[0,1]$.}}
\label{Fig4}
\end{figure}

In this case, both types of states are polynomials in $\gamma$ and only even combinations of eigenfunctions appear (see \eqref{vacuumho} and \eqref{vacuume}). Again, the energy-like states have more stable trajectories but for small values of $\gamma$. In Fig \ref{Fig4}, we have compared  the trajectories for the energy-like states when $\gamma=0.2, 0.5$ and $0.7$.

At $t=0$, the uncertainty product $\Delta(0, \gamma)=(\Delta x)^2 (\Delta p)^2$ and the dispersion $(\Delta x)^2$ take almost the same values for both types of states. Minimal uncertainty is satisfied for $\gamma\in [0,0.2]$ and we get very small values of the dispersion in $x$ for the same values of $\gamma$. The behaviour is similar to the one in Fig \ref{Fig2}.

\begin{figure}[!h]
\centering
\includegraphics[width=.5\textwidth]{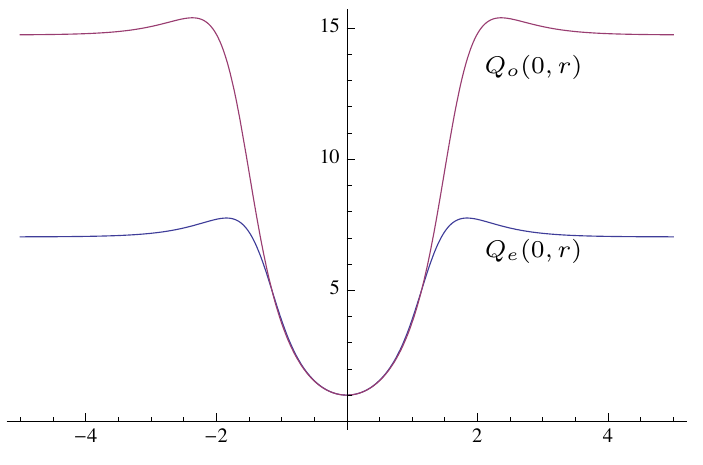}
\caption{ {\small Comparison of Mandel parameter $Q(0,\gamma)$ in the vacuum for the energy-like and oscillator-like squeezed states as a function of $r$ such that $\gamma=\tanh r$.}}
\label{Fig5}
\end{figure}

\begin{figure}[!b]
\centering
\includegraphics[width=.45\textwidth]{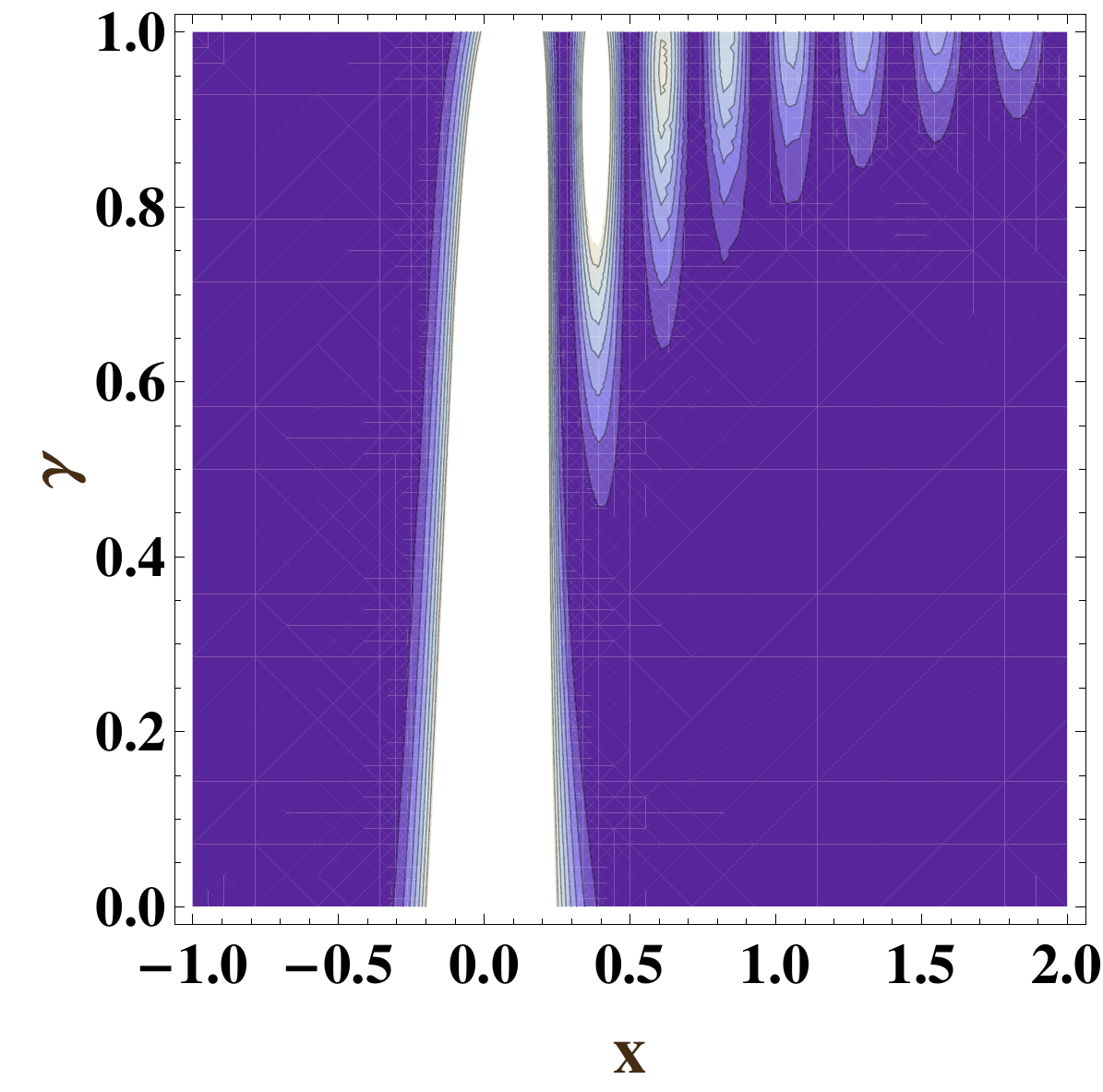}\qquad \qquad
\includegraphics[width=.45\textwidth]{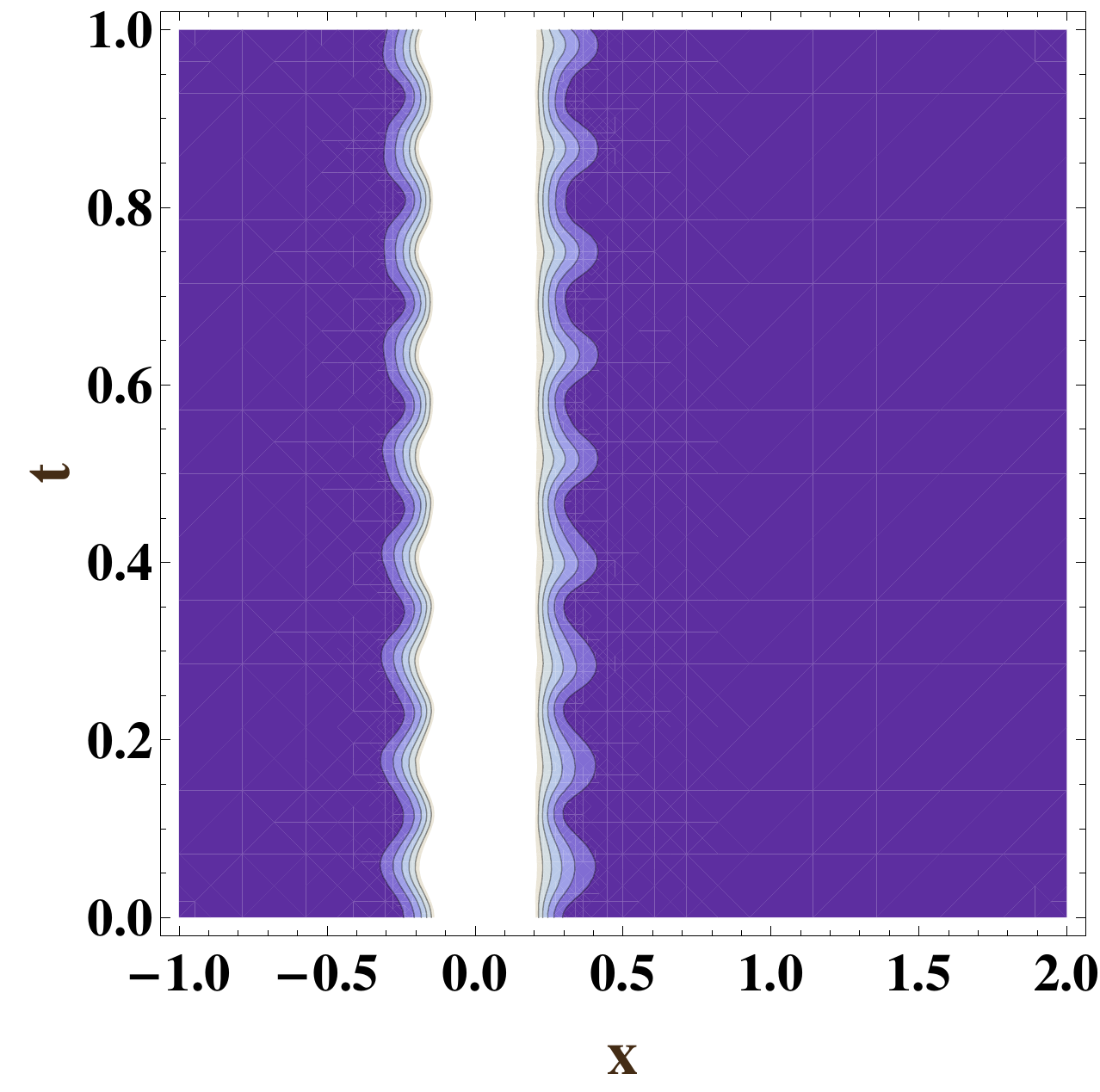}
\caption{ {\small Density probability $|\Psi^{\nu}_e(0, \gamma,x;t)|^2$ in the vacuum for energy-like coherent states for $x\in[-1,2]$ as a function of $\gamma \in [0,1]$ with $t=0$ (left) and a function of $t \in [0,1]$ with $\gamma =0.2$ (right).}}
\label{Fig6}
\end{figure}

With respect to the statistical properties, Fig  \ref{Fig5} shows bunching effects of $Q(0, \gamma)>0$ both for the energy-like and oscillator-like states. It confirms that both types of states have similar statistical behaviour.   
Finally, the density probability has been computed for the energy-like states. In Fig \ref{Fig6} we see that the best localisation occurs when $\gamma$ is smaller than $0.2$. We see also, by observing the time evolution, that the vacuum states are less stable than the coherent states.
Similar behaviour is observed for the oscillator-like states.

\subsection{General system of states}

\begin{figure}[!b]
\centering
\includegraphics[width=.5\textwidth]{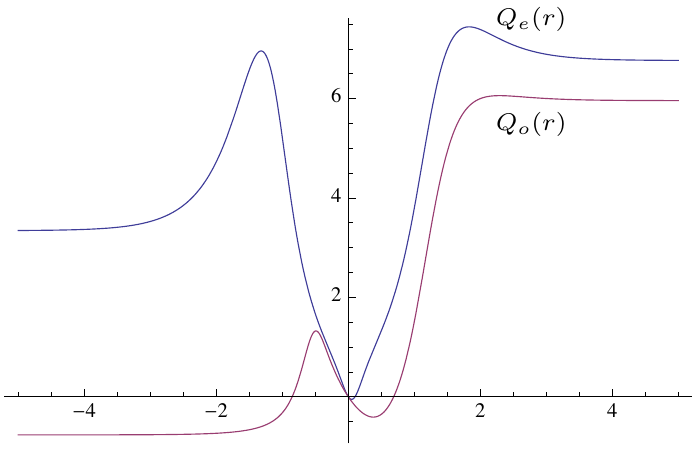}
\caption{ {\small Comparison between the Mandel parameter for the energy-like and oscillator-like squeezed states as a function of $r$ such that $\gamma=\tanh r$ for $z=2$.}}
\label{Fig7}
\end{figure}

In this case, we observe that the trajectories deviate quickly from closed curves while the minimum uncertainty is still realised but when $\gamma < 0.2$ .
The graphs for the Mandel's parameter for $z=2$ are given in Fig \ref{Fig7}. The bunching  and anti-bunching  are observed for both types of states,  the bunching is more prominent for energy-like states for all values of $r$.  More significant anti-bunching effect is observed for oscillator-like states showing a steady effect for  $r<0$ and a minimum for  $r>0$ (when $\gamma \approx 0.2$).

The calculations of observables for other diatomic molecules can be done in a similar way.  For example, we have done such calculations for the molecule $^{133}{\rm Cs}_2$ which has a larger value for $\nu$. Indeed, we have $\nu\approx 524.55$ and thus $[p]=261$. We have obtained similar behaviour for both types of states that is not relevant to produce here.

\section {Conclusions \label{sec:conclusions}}

In this paper, we have introduced squeezed coherent states of a quantum system with a finite discrete energy spectrum described by the Morse potential.
These states are almost eigenstates of linear combination of ladder operators and are characterised by two continuous parameters $z$ and $\gamma$. 
We have considered two different types of ladder operators and constructed the corresponding oscillator-like and energy-like squeezed coherent states.

We have investigated the behaviour of these states regarding localisation and minimum uncertainty. The calculation of the dispersions and mean values have been done analytically except for the mean value of $\hat{x}^2$ for which the analytical form is not known. We have computed the Mandel's parameter to investigate the statistical properties of our states.

The oscillator-like squeezed coherent states are closely related to the similar states of the harmonic oscillator. However, they do not have a very good localisation except for the vacuum case for $\gamma$ small and they exhibit a certain deviation from the minimal uncertainty principle. The energy-like squeezed coherent states minimise better the uncertainty relation for the case $\gamma=0$ and we get a good localisation in position.
They are more stable in time than the oscillator-like states. Both type of states have similar statistical properties.

\section*{Acknowledgements}

V. Hussin acknowledge the support of research grants from NSERC of Canada. A. Hertz acknowledge the support of a NSERC research fellowship. This work has been started while V. Hussin visited Northumbria University (as visiting professor and sabbatical leave). This institution is acknowledged for hospitality. The authors thank J. Van der Jeugt for helpful discussions on special functions and orthogonal polynomials.


\section*{References}
\begin{thebibliography}{99}

\bibitem {schro 26} Schr\"{o}dinger E, 1926 {\it Naturwiss} {\bf 14} 664.

\bibitem {kenn 27} Kennard E H 1927 {\it Zeit Phys} {\bf 44} 326.

\bibitem{Glauber} Glauber R J, 1963 {\it Phys Rev} {\bf 130} 2529;  {\bf 131} 2766.

\bibitem{Klauder} Klauder J R , 1960 {\it Ann Phys} {\bf 11} 123.

\bibitem{KS} Klauder J R and Skagerstam B S, 1985
{\it Coherent States-Applications in Physics and
Mathematical Physics} (World Scientific, Singapore).

\bibitem{Nieto} Nieto M M,1997, arXiv: quant-ph/9708012.

\bibitem{Walls} Walls D F and Milburn G J 2008 {\it Quantum Optics 2nd Edition} (Springer, Berlin).

\bibitem {Gazeau} Gazeau J P 2009 {\it Coherent states in Quantum Physics}
(Wiley, New York).

\bibitem {Rand} Rand S C 2010 {\it Nonlinear and Quantum Optics}
(Oxford University Press, Oxford).

\bibitem{Braunstein} Braunstein S L and  McLachlan R I 1987 {\it Phys Rev A} {\bf 35} 1659-1667.

\bibitem{Hillery} Hillery M 1987 {\it Phys Rev A} {\bf 36} 3796-3802.

\bibitem{Bergou} Bergou J A, Hillery M and Yu D 1991  {\it Phys Rev A} {\bf 43}, 515-520.

\bibitem{Sasaki1} Fu H-C and Sasaki R 1996  {\it Phys. Rev. A} {\bf 53}, 3836-3844.

\bibitem{Alvarez} Alvarez N and Hussin V 2002  {\it J Math Phys} {\bf 43} 2063-2085.

\bibitem{Klauder01} Klauder J R, Penson K A , and Sixderniers J M 2001 {\it Phys. Rev. A} {\bf 64}013817.

\bibitem{Dong1} Dong S H 2002 {\it Can. J. Phys.} {\bf 80} 129-139.

\bibitem{Roy} Roy B and Roy P 2002 {\it Phys. Lett. A} {\bf 296} 187-191.

\bibitem{Recamier}{R\'ecamier J and J\`auregui R} 2003 {\it J Opt B} {\bf 5} S365-S370.

\bibitem{Daoud} Daoud M and Popov D 2004 {\it Int J Mod Phys B} {\bf 18} 325-336.

\bibitem{Angelova} Angelova M and Hussin V 2008 {\it J Phys A} {\bf 41} 30416.

\bibitem{Frank} Dong S H, Lemus R and Frank A 2002 {\it Int J Quant Chem}  {\bf 86}
433.

\bibitem{Dong2} Dong S-H 2008 {\it Factorization Method in Quantum Mechanics, Fundamental theories in physics 150} (Springer, Dortrecht, The Netherlands).

\bibitem{Fox} Fox  R F and Choi M F 2001 {\it Phys. Rev. A} {\bf 64} 042104.

\bibitem{Draganescu} Draganescu G E, Messina A and Napoli A 2009 {\it J Mod Optics} {\bf 56} 508-515.

\bibitem{Yuen} Yuen H P 1976 {\it Phys. Rev. A} {\bf 13} 2226.

\bibitem{Szafraniec} Szafraniec F H 1998 {\it Contemp. Math.} {\bf 212} 269-276.

\bibitem{Merzbacher}  Merzbacher E 1998{\it Quantum Mechanics} (Wiley, New York).

\bibitem{Singh} Singh A C and Babynanda D O 2006, {\it Int J Quantum Chemistry} {\bf 106} 415-425.

\bibitem{Sasaki} Odake S and Sasaki R 2006 {\it J Math Phys} {\bf 47} 102102.

\bibitem{Sage} Sage M L 1978  {\it Chemical Physics} {\bf 35} 375.

\bibitem{Mandel} Mandel L 1979 {\it Opt Lett} {\bf 4} 205.

\bibitem{Koe} Koekoek R and Swarttouw R F 1998 {\it The Askey-scheme of hypergeometric orthogonal polynomials and its $q$-analogue} (Technical report 98-17, Delft University of Technology), http:// fa.its.tudelf.nl/koekoek/askey/contents.html.

\bibitem{Herzberg50} Herzberg G 1950 {\it Molecular Spectra and
Molecular structure} Vol. I: Spectra of Diatomic Molecules, 2nd
edition (Van Nostrand, Princeton).

\bibitem{Guela} Guelachvili G, Noah P and Bedace P 1981 {\it J Mol Spectr} {\bf 85} 271-281.

\bibitem{CRC} {\it CRC Handbook of Chemistry and Physics} 90th Edition 2009-2010,
Ed. David R. Lide, Section 9, Molecular Structure and Spectroscopy (CRC netBase, 2010 Taylor and Francis).

\bibitem{AF07} Angelova M and Frank A 2005 {\it Phys At Nuclei} {\bf 68}
1625.

\end{thebibliography}
\end{document}